\documentclass[conference]{IEEEtran}

\makeatletter\def\input@path{{current}}\makeatother

\usepackage{amsmath,amssymb,amsfonts,amsthm}
\usepackage{algorithmic}
\usepackage{graphicx}
\usepackage{textcomp}
\usepackage{xcolor}
\usepackage[utf8]{inputenc}
\usepackage[margin=1in]{geometry}
\usepackage[T1]{fontenc}

\usepackage[english]{babel}
\usepackage[square, numbers]{natbib}                                                   
\usepackage{xfrac}

\usepackage[nolist]{acronym}

\def\BibTeX{{\rm B\kern-.05em{\sc i\kern-.025em b}\kern-.08em
    T\kern-.1667em\lower.7ex\hbox{E}\kern-.125emX}}

\newcommand{\blue}[1]{#1} % turn off blue highlighting

\newcommand\bonware{\mathcal{B}}
\newcommand\malware{\mathcal{M}}

\newcommand\ma{\mathcal{A}} % mission accomplishments

\newcommand\Fnominal{\blue{F_\text{N}}{}}

\IEEEoverridecommandlockouts
\begin{document}

\graphicspath{ {./figures/} }

\title{An Experimentation Infrastructure for
  Quantitative Measurements of Cyber Resilience\thanks{This work
    was partially funded by Cyber Technologies, Deputy CTO for
    Critical Technologies/Applied Technology, Office of the Under
    Secretary of Defense Research and Engineering. Contributions from 
    Dr. Vandekerckhove were accomplished under Cooperative Agreement 
    Number W911NF-21-2-0284 with ARL. Mr. Ellis completed this work while
    a contractor with ICF International, and has since transitioned to a 
    civilian role within ARL.}}

\newcommand\nobody{\IEEEauthorblockN{\phantom{nobody}\\\phantom{nowhere}\\}}

\newcommand\arl{\IEEEauthorblockA{$^{1}$%\IEEEauthorrefmark{1}
    \textit{U.S. Army Combat Capabilities Development Command}\\
    \textit{Army Research Laboratory}, Adelphi, MD\\
    \{jason.e.ellis7,sidney.c.smith24,alexander.kott1,michael.j.weisman2\}.civ@army.mil}}

\newcommand\icf{\IEEEauthorblockA{$^{2}$%\IEEEauthorrefmark{2}
    \textit{ICF International}, Aberdeen, MD \\
    travis.w.parker16.ctr@army.mil}}

\newcommand\uci{\IEEEauthorblockA{$^{3}$%\IEEEauthorrefmark{3}
    \textit{University of California, Irvine} \\
    \textit{Department of Cognitive Sciences}, Irvine, CA \\
    joachim@uci.edu}}

\newcommand\psu{\IEEEauthorblockA{$^{4}$%\IEEEauthorrefmark{4}
    \textit{Pennsylvania State University}\\
    \textit{Applied Research Laboratory}, University Park, PA \\
    bjm206@arl.psu.edu}}

\newcommand\names{\IEEEauthorblockN{Jason E. Ellis$^{1}$, Travis W. Parker$^{2}$, Joachim Vandekerckhove$^{3}$,
    Brian J. Murphy$^{4}$,\\ Sidney Smith$^{1}$, Alexander Kott$^{1}$, Michael J. Weisman$^{1}$}}

\newcommand\affiliations{\arl \icf \uci \psu}

% \author{\arl \and \nobody \and \icf \and \uci \and \psu}
\author{\names \affiliations}

\maketitle

% \tableofcontents

\begin{abstract}
The vulnerability of cyber-physical systems to cyber attack is well
known,
and the requirement to build cyber resilience into these systems has
been firmly established.
The key challenge this paper addresses is that maturing this discipline
requires the development of techniques, tools, and processes for
objectively, rigorously, and quantitatively measuring the attributes of
cyber resilience.
Researchers and program managers need to be able to determine if the
implementation of a resilience solution actually increases the
resilience of the system.
In previous work, a table top exercise was conducted using a notional
heavy vehicle on a fictitious military mission while under a cyber attack.
While this exercise provided some useful data, more and higher
fidelity data is required to refine the measurement methodology.
This paper details the efforts made to construct a cost-effective
experimentation infrastructure to provide such data.
It also presents a case study using some of the data generated by the
infrastructure.
\end{abstract}
 
\section{Introduction}
Researchers have demonstrated the vulnerabilities of modern vehicles to
cyber attacks \cite{
	hoppe2007sniffing,
	koscher2010experimental,
	bozdal2018hardware,
	bozdal2018survey,
	stachowski2019cybersecurity,
	bozdal2020evaluation}.
Many realize that perfect security is impractical, and that resilience is
necessary \cite{
	obama2011ppd-8,
	obama2013ppd-21,
	alexeev2017constructing,
    % henshel,
	linkov2018risk,
	kott2019cyber}.
In order for the field of cyber resilience to mature,
there must be "techniques, tools, and processes for objectively
measuring the attributes of phenomena occurring in the systems of
that discipline'' \cite{kott2021improve}.

Here, we ask if we can construct an inexpensive experimental
environment that produces data of sufficient quality to demonstrate a
quantitative measurement of cyber resilience.
The development of a cyber resilience
measurement methodology does not require high fidelity experimental
environments at this time;
it is enough to reasonably approximate the performance of a generic
\ac{MGV}.

This paper describes an inexpensive experimental environment that may be 
used to test various malware and bonware (the totality of physical and cyber features that allow a system to resist and recover from cyber compromise) on the \ac{CAN} bus of a modern vehicle.
A case study is presented, and the data from that case study is used to 
measure the cyber resilience of the simulated vehicle. 

% The development of a quantitative measure of cyber resilience begins
% with a clear definition of cyber resilience.
In Section~\ref{sec:prior_work}, we begin by reviewing the categories of methods
used in researching vehicle cybersecurity.
The various components of the experimental infrastructure are described in Section~\ref{sec:components}.
An illustrative case study is presented in Section~\ref{sec:case_study}.
In Section~\ref{sec:conclusion}, we provide discussion and concluding
remarks.

\section{Prior Work}\label{sec:prior_work}
Much of the early work on the cybersecurity of automobiles was done using
actual vehicles \cite{
	hoppe2007sniffing,
	koscher2010experimental,
	miller2013adventures,
	miller2015remote,
	foster2015fast}.
This provides the highest fidelity; however, it is very expensive to conduct
research in this fashion.
The number of runs required to generate the data necessary to validate a 
quantitative measurement of cyber resilience would be prohibitively
expensive. 

Some researchers conducted research by connecting multiple \acp{ECU} together
on a \ac{CAN} bus independent of a
vehicle \cite{
	daily2016towards,
	bozdal2018hardware,
	wang2018delay}.
These do indeed provided an inexpensive method to test malware and bonware
in a vehicular network; however, they cannot capture impacts on the 
vehicle's \acp{KPP}.

Shikata et al.~ \cite{shikata2019digital} developed a Digital Twin:
a system to reproduce real-world events in a digital
environment.
A virtualized vehicle with realistic virtual performance would provide
high fidelity at low cost in terms of time to test and measure cyber resilience.
However, constructing a virtual vehicle would be prohibitively expensive
unless it was constructed by the vehicle manufacturer to facilitate design.

\section{Component Descriptions}\label{sec:components}
\begin{figure}[htbp]
	\centerline{\includegraphics[trim=0 10 0 15,scale=.95]{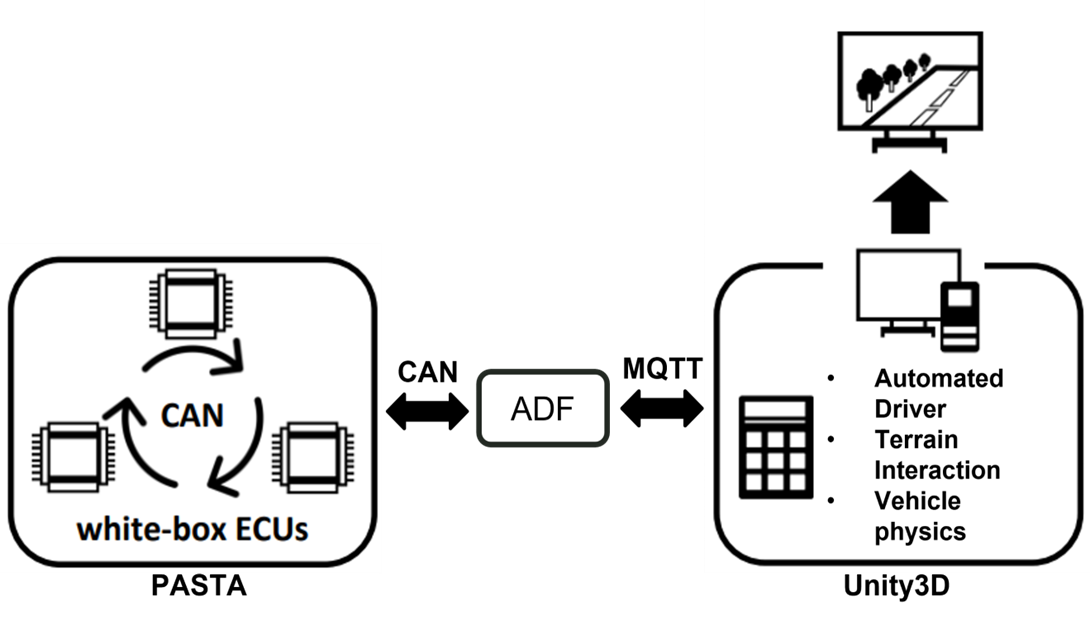}}
	\caption{
		A high-level overview of the data flow between components.
		Portions are derived from \cite{toyama2018pasta}.
	}
	\label{fg:datamodel}
\end{figure}

Multiple technologies have been integrated together to facilitate the
generation of vehicle data.
In our experimentation setup, these include the \ac{PASTA} by Toyota Motor Corporation,
the Unity game development platform, the \ac{ADF} developed in-house here
at the DEVCOM Army Research Laboratory,
and the OpenTAP test automation framework by Keysight Technologies.
In general, Unity generates messages via the \ac{MQTT} 
publish-subscribe network protocol.
\ac{ADF} ingests these messages and translates them to \ac{CAN} format,
which are then injected onto the appropriate \ac{CAN} bus within \ac{PASTA}.
Data flowing in the opposite direction is handled in a similar manner,
but in reverse. Figure~\ref{fg:datamodel} illustrates the flow of data between
components.

\subsection{The \ac{PASTA} Testbed}
\ac{PASTA}, created by Toyota, is a vehicle security testbed designed to
allow researchers the ability to develop and evaluate new security technology
and approaches on realistic ``white-box'' \acp{ECU} \cite{toyama2018pasta}.
There are three vehicle \acp{ECU} within the testbed, each encompassing a
family of functionality and sitting on their own \ac{CAN} bus: powertrain,
body, and chassis.
These three \acp{ECU} are responsible for their respective group of messages,
each generating and responding to traffic on their bus.
A fourth \ac{ECU}, the \ac{CGW}, acts as a junction between the three
previously mentioned buses.
Based on the message and source bus, the \ac{CGW} ferries messages to their
appropriate destination bus.
The firmware for all \acp{ECU} is open-source, and is accompanied by an
\ac{IDE} from Toyota to promote ease of customization.

\ac{PASTA} has integrated simulation boards which calculate
how the current \ac{CAN} messages on the buses would physically influence a commercial sedan.
These boards then update the vehicle \acp{ECU} with appropriate values.
For example, when acceleration pedal operation is inputted, the chassis
\ac{ECU} sends a message with the indicated value.
The simulation boards observe this message and calculate the physics that
would result from the input.
The results are used to update the values reported by the powertrain \ac{ECU},
which it outputs onto its bus.
In this instance, the powertrain ECU would send messages indicating the new
engine throttle position, \ac{RPM}, and vehicle speed.
The simulation boards are connected directly to the vehicle \acp{ECU} via a
serial port on each ECU board, so the physics data is not mixed with messages
on the \ac{CAN} bus.

An issue with the simulation boards is that they are essentially
``black-boxes'' and closed-source.
We cannot alter, for instance, the parameters involving the engine (e.g.,
torque and horsepower), the weight of the vehicle, or the terrain that the
boards are assuming is being traversed.
To counteract this issue, we integrated a simulation engine that would allow
for user-defined vehicle details as well as a custom terrain.
In this configuration, \ac{PASTA} becomes hardware-in-the-loop for the
simulation engine.
Attacks or defenses that affect the \acp{ECU} present in \ac{PASTA} will also
affect the vehicle within simulation.
This is essential to observe realistic threats and defenses and
their effects on vehicle performance.

\subsection{Unity}
Unity is a widely used multi-platform game development platform implemented
in C\# \cite{unity2021unity}.
It includes out-of-the-box tools, assets, and an \ac{IDE} to aid developers
in creating 3D designs and models.
In particular, Unity provides built-in assets and classes regarding vehicle
physics, which we leverage to model interactions between our simulated
vehicle and custom terrain.

\subsubsection{Simulated Light Military Ground Vehicle}
We implemented a light \ac{MGV} within Unity, inspired by the \ac{JLTV}, designed to interface
with data inputs from the white-box \acp{ECU} within \ac{PASTA}.
In general, the vehicle produces inputs in response to the simulated terrain.
These inputs are sent to the corresponding \acp{ECU} within \ac{PASTA} as if
they were generated by \ac{PASTA} itself.
We then gather responses to these inputs from the \acp{ECU} and send them back to the simulated vehicle, which it uses to calculate parameters like torque and fuel
consumption.
For example, assume that the vehicle reports that the accelerator is set to
50\%.
This message is injected into \ac{PASTA} as if the chassis \ac{ECU} had
generated it.
The powertrain ECU responds to the message with the corresponding amount of
engine throttle.
A message with the engine throttle is sent back to Unity, which is then
applied to engine power calculations.
With this flow, any attacks present on the \acp{ECU} within \ac{PASTA}
will affect the simulated vehicle.

An automated driver is responsible for generating steering, acceleration,
and braking inputs as the vehicle traverses the simulated terrain.
Steering is guided by the use of a waypoint system.
Both acceleration and braking inputs are calculated via a
\ac{PID} controller.
The controller reliably responds to sudden changes in the terrain or vehicle
performance, and maintains the target vehicle speed while preventing
speed oscillation.

Engine performance is calculated through the use of torque, horsepower, and
\ac{BSFC} curves.
Engine RPM is derived using the vehicle speed, wheel circumference, and
effective gear ratio.
Using this \ac{RPM} value, the curves are evaluated to discern the
corresponding torque, horsepower, and \ac{BSFC} value.
Torque is multiplied by the throttle and the current effective gear ratio to
obtain the total amount of torque that can be applied to the wheels.
Since our vehicle is  \ac{AWD}, each wheel receives the total amount of
torque divided by the number of wheels on the vehicle, which is six in this
case.
Horsepower and the \ac{BSFC} value are used in conjunction to calculate the
amount of fuel that has been used.
Based upon the torque curve as a whole, gear shift points were calculated
manually.
The resulting shift points, being optimal in terms of torque output, confer
an aggressive acceleration profile for the vehicle.
% Appendix A
% contains formulas used to calculate the engine \ac{RPM}, the total current
% torque, and fuel consumption at each update of the physics engine.

The vehicle is capable of providing sensor information that is either not
present in \ac{PASTA} or needs its functionality altered for our purposes.
Currently, this applies to the engine coolant temperature.
Engine temperature is present in \ac{PASTA}, but is aligned to the
temperature characteristics of a commercial sedan.
Within Unity, we implemented a basic temperature model that can be controlled
by an external fan controller \ac{ECU} described in a later section.
The fan controller monitors the coolant temperature reported by the vehicle and dictates the operation of a simulated fan on the vehicle.
% Once it reaches an upper threshold limit, it activates the simulated fan on
% the vehicle, which begins to lower the temperature to an acceptable lower
% threshold.
The fan itself takes around 50 horsepower to operate, which results in an
approximate 25\% drop in the available torque that can be applied to the
wheels.

\subsubsection{Terrain}
The simulated vehicle within Unity traverses a custom terrain map that is
roughly 81.8 km by 100 km with altitudes up to 910 m.
We crafted a course approximately 151 km in length across the map that
encompasses multiple terrain types: flat main road, float off-road, hilly,
prolonged incline, and prolonged decline.
We placed trigger points on the map that set the target speed of the vehicle
depending on the terrain type.
If on a flat main road, the target speed is 60 km per hour.
Otherwise, the target speed is 40 km per hour.

\subsection{Active Defense Framework}
\ac{ADF} is a government-developed framework used to rapidly prototype
network-based active cyber-defense techniques.
\ac{ADF} currently supports \ac{IP} networks and vehicle control networks,
namely the \ac{CAN} bus and \ac{SAE} J1708 bus.
The framework itself is written in Python and C programming languages,
while active cyber-defense plugins can be written in Python.
The framework acts as an intermediary for network traffic, as depicted in
Figure~\ref{fg:datamodel}, allowing it to control network message flow, as well
as inspect, modify, drop, or generate network messages.
Essentially, the plugin developer has free reign to prototype any
network-based cyber-defense or detection technique they can imagine.

% \ac{ADF} is a fully modular architecture comprising the core framework and
% plugins.
% The main framework process is responsible for managing plugins and handling
% events, but it does not play any role in packet processing, and it can handle
% any type of packet- or message-based traffic. 

% Each plugin is a self-contained module.
% Plugins are linked by the framework based on the configuration.
% Packets arrive at a plugin with information (a dictionary of attributes)
% generated by the previous plugins.
% After taking some action on the packet or the information, the packet is
% dispatched to other plugin(s) based on the configured links and the packet
% information.
% Specialized plugins are used to send and receive traffic via network
% interfaces, buses, and other ADF instances.
% By creating the proper plugins and linking them, ADF can act as a firewall,
% bridge, gateway, or honeypot.

% Each plugin can maintain state data.
% The state of all plugins can be automatically saved and loaded by the
% framework to provide persistence. 

% Plugins communicate with each other by generating events, which are placed
% into the framework’s event queue.
% The framework pulls events from this queue and notifies the plugins in a
% configurable order.
% Each plugin can contain logic to modify the event or act based on it.

\ac{ADF} is used here to provide a system/software-in-the-loop capability,
and to provide simulated components.
It allows communication between \ac{PASTA} and Unity by translating \ac{CAN}
messages to and from \ac{MQTT}, a standard publish-subscribe \ac{IP}-based messaging protocol.
\ac{ADF} plugins are also used to provide simulated \ac{ECU} hardware, and to
implement attack and defense methods on the CAN bus via ADF’s ability to
monitor, modify, inject, or drop CAN traffic.

\subsubsection{Unity-to-PASTA Message Translation}
\ac{ADF} runs on a standalone laptop and is connected to \ac{PASTA} via two
\ac{USB} \ac{SLCAN} interface modules.
One module is connected to the powertrain \ac{CAN} bus, and the other
module is connected to the chassis CAN bus.
The \ac{PASTA} \ac{CGW} is disconnected from the CAN buses for our
experiments, and the body \ac{CAN} bus and body \ac{ECU} are not
used.
\ac{ADF} is configured to serve as a \ac{CGW} between Unity and \ac{PASTA}. Since Unity does not natively communicate with \ac{CAN} interfaces, \ac{ADF} translates \ac{CAN} messages in real-time to \ac{MQTT} messages and back. Unlike the \ac{PASTA} \ac{CGW}, \ac{ADF} does not relay messages between the powertrain and chassis \ac{CAN} buses themselves.
\ac{ADF} relays powertrain \ac{CAN} messages between Unity and \ac{PASTA}, and sends
vehicle parameters from Unity to the chassis \ac{CAN} bus for display on the
\ac{PASTA} instrument cluster. All communication channels are two-way.

\subsubsection{Virtual \acp{ECU} within \ac{ADF}}
The \ac{PASTA} platform does not simulate a controllable cooling fan or
provide a fan controller \ac{ECU}.
Therefore, we simulate a fan controller \ac{ECU} using an \ac{ADF} plugin.
The fan controller engages the engine cooling fan on the simulated vehicle when the engine coolant temperature reaches a defined upper limit, and disengages the fan
when temperature drops below a lower limit.
For the purposes of our experiments, the fan controller \ac{ECU} plugin can
simulate an attack on its own firmware, stop the simulated attack, or reset/``re-flash'' itself (i.e., replace the \ac{ECU} firmware). During a reset, the fan controller is offline for 20~s.

We can create other simulated \acp{ECU} using \ac{ADF}. For instance, a \ac{CTIS} module is not implemented within \ac{PASTA}. This functionality could be captured and controlled by a plugin to extend the attack surface of the simulated vehicle.

\subsubsection{Attack Possibilities via \ac{ADF}}
The simplest attack on a vehicle bus is to inject messages.
Messages are broadcast on a \ac{CAN} bus, so one injected at any point
on the bus will reach all \acp{ECU} on the bus.
While injection attacks cannot block or modify normal \ac{CAN} bus
traffic, they can impact vehicle performance if we inject messages to
cause constant undesired vehicle behavior. 
% For example, in heavy vehicles a \ac{PGN} 0 message is emitted by an \ac{ECU}
% that needs to control the engine speed or output torque.
% Normal sources of \ac{PGN} 0 include the brake controllers when loss of
% traction is detected, and the automatic transmission controller when the
% engine torque/speed needs to be reduced to safely shift gear.
% If a \ac{PGN} 0 message demanding idle speed or torque is continually
% injected on the bus, the engine will be forced to idle and the accelerator
% will be unresponsive.

If an attacker can physically sever the \ac{CAN} bus wiring at a strategic
point and place additional hardware there, it is possible to block or modify the normal bus traffic.
Attacks that block or modify messages can prevent \acp{ECU} from controlling
the vehicle or falsify vehicle data. % One example would be reporting a lower
% than actual coolant temperature, preventing the fan from engaging and
% eventually causing the engine to overheat.
% Another example would be blocking legitimate \ac{PGN} 0 control messages,
% which would disable traction/stability control or prevent the transmission
% from shifting.

As a man-in-the-middle between \ac*{PASTA} and Unity, \ac*{ADF} can execute any of these bus-level attack types.

Attacks on \ac{ECU} firmware, by embedding malware, are also feasible.
Malware on the fan controller is simulated by sending a trigger event to the \ac{ADF} plugin.
% The attack we simulated is a ``stuck fan'' attack in which malware has
% modified the fan control \ac{ECU} to not disengage the fan once engaged,
% even when the coolant temperature has dropped below the minimum operational
% temperature.
% This does not result in an immediately noticeable condition such as
% overheating, but will reduce powertrain output and increase fuel consumption.

\subsubsection{Defensive Capabilities via \ac{ADF}}
Defending against message injection, blocking, and modification at the bus-level requires detecting and filtering injected messages before they reach the \ac{ECU}.
The \ac{CAN} bus can be split at potential access points and hardware placed
in-line, hardware can be placed between the \ac{CAN} bus and critical
\acp{ECU}, or defenses can be integrated into the \acp{ECU} themselves. Examples of these defenses implemented previously using \ac{ADF} include cryptographic watermarking and modeling observable vehicle states to compare current parameters to the model's prediction.

% Detection of injected or modified messages can be achieved by watermarking
% valid messages.
% A cryptographic watermark can be inserted into the reserved or
% least-significant bits of a message before it traverses the bus and verified
% by recipients.
% Alternatively, messages can be cryptographically signed, and the signature
% transmitted as an additional message.
% Messages injected or modified by the attacker will not have a valid watermark
% or signature and will be rejected. 

% Message blocking or modification can also be detected by modeling the vehicle
% based on observable state and comparing the parameters in the suspect message
% to the model’s prediction.
% Additionally, if messages are being blocked, or messages with invalid
% watermarks/signatures are rejected, those messages need to be generated.
% For example, coolant temperature can be modeled based on intake air
% temperature, fuel temperature, and oil temperature.
% If the coolant temperature message is missing or not within range of the
% prediction, we can substitute a message containing the predicted coolant
% temperature. 

Attacks on \acp{ECU} themselves must be approached differently.
If an \ac{ECU} is compromised, measures need be taken to restore proper
\ac{ECU} function.
Many \acp{ECU} can be re-flashed while the vehicle is operational.
The \ac{ECU} may or may not be functional for some duration while being reset
or re-flashed, and the impact this will have to vehicle performance depends
on the function of the \ac{ECU}.
For the \acp{ECU} simulated by ADF plugins, the behavior is to make the 
\ac{ECU} unresponsive for a set duration, after which normal \ac{ECU}
operation is restored.

\subsection{OpenTAP}
OpenTAP is an open-source test automation framework developed by Keysight Technologies \cite{opentap2021whitepaper}. It provides a test sequencer to promote test repeatability, a customizable plugin facility capable of integrating plugin classes implemented in C\# or Python, and result listeners responsible for capturing test data for further analysis. OpenTAP is used to automate the execution of experiments and provide a GUI for testing practitioners to configure experiment steps.

\subsection{Reportable Data}
Data can be captured from two different yet related perspectives: the simulated vehicle or the PASTA CAN buses. Both views have their respective formats: a CSV containing multiple columns of data delineated by timestamp from the vehicle, or a CAN-formatted capture of traffic from the PASTA buses. % Appendix B contains a table showing the data reported by the vehicle, where \cite{pasta2019canids} extensively lists the CAN messages seen from PASTA.

Within our infrastructure, there are multiple parameters that can be configured to generate varied data captures. Currently, these include experiment duration, attack start time, terrain type(s), starting location, ending location, target vehicle speed, and attack-defense pairing.

\newcommand{\kpp}{\mathcal K}
\section{Illustrative Case Study}\label{sec:case_study}
\subsection{Scenario Description}
We created and ran an experiment scenario wherein malware has infected
the fan controller \ac{ECU} present on the vehicle.
The malware, once active, removes the ability to disengage the engine coolant
fan from the fan controller \ac{ECU}.
This causes the vehicle to enter a state of prolonged degraded performance.
% due to the fact that the engine coolant fan draws approximately 25\% of the
% available horsepower in order to operate.
Note that we did not employ actual malware in this experiment.
We simulated the effects of such a malware and applied them to the fan
controller \ac{ECU} implemented within \ac{ADF}.

The experiment runs for $800~s$. %with the attack commencing around
% 450 seconds from the start of the run.
Simulated bonware present on the vehicle monitors the engine coolant temperature for any abnormalities.
From this sensor reading, it will notice the effect of the malware and
attempt to recover after a period observation to ensure the behavior
is not a short-lived anomaly.
Recovery from this attack consists of re-flashing the firmware on the fan
controller \ac{ECU}.
% On an actual \ac{ECU}, this process takes approximately 20 seconds to
% complete.
% Since our fan controller is implemented within \ac{ADF}, this process is also
% simulated.

\begin{figure*}[bth]
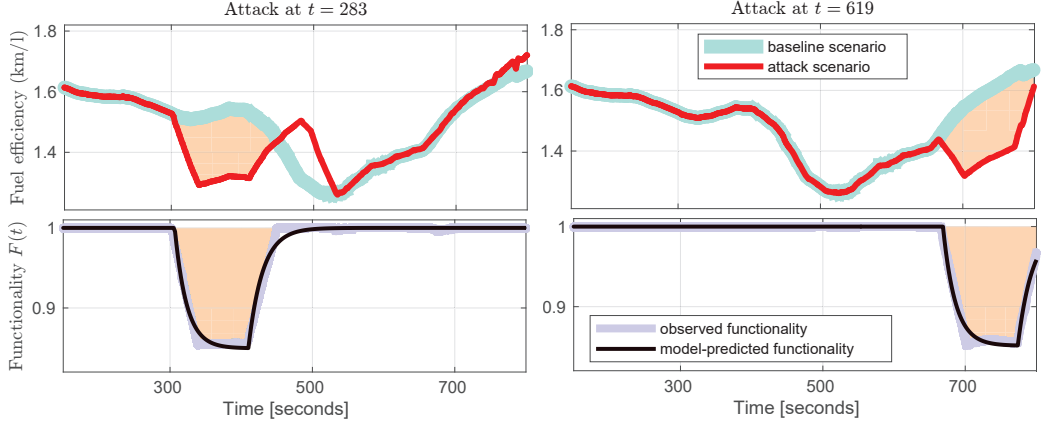


    \newcommand{\myig}[1]{\includegraphics[scale=.75]{#1}}
    
    \centering
  
    \myig{piecewise_ode_with_malware_lag-283-fuelEfficiency-milcom2-data}
    \myig{piecewise_ode_with_malware_lag-619-fuelEfficiency-milcom2-data}\\
    \myig{piecewise_ode_with_malware_lag-283-fuelEfficiency-milcom2-fcn}
    \myig{piecewise_ode_with_malware_lag-619-fuelEfficiency-milcom2-fcn}
  
    \caption{%
      \textbf{Top row}: Fuel efficiency curves under baseline (cyan) and attack (red) scenarios, with the effect of malware shaded orange.
      \textbf{Bottom row}: Observed (blue) and inferred (black) functionality curves for the time series model. A change in the timing of the attack ($t = 283$ vs,\ $t = 619$) results in a simple translation of the functionality curve.
    }
  
    \label{fig:fuel-efficiency}
  
\end{figure*}
\subsection{Quantitative Measures of Cyber Resilience}
In \cite{mmcr} we develop a model for the behavior of a
system's functionality over the course of an incident where it is
being attacked by malware and defended by bonware.  First, we assume
that there is an observable, sufficiently smooth function representing mission
accomplishment, and we define functionality to be its time
derivative.  Thus,
\begin{equation*}
  \label{eq:maf}
  F(t) = \frac{d \ma }{dt}, \quad \ma(t) = \int_{t_0}^t F(\tau) \, d\tau.
\end{equation*}
The normal
functionality, when the system performs normally and does not
suffer a cyber attack, may, in general, vary with time.
For simplicity, throughout the paper, we assume the normal
functionality to be constant in time, $F_0(t)=\Fnominal.$ 

To be able to compare measures of cyber resilience for multiple
missions, we normalize the measure by dividing by the total mission
time.
\begin{equation*}
  R= \frac{1}{T-t_0} \int_{t_0}^T F(\tau) \, d\tau =
  \frac{\ma(T)}{T-t_0}.
\end{equation*}
Often there are multiple objectives to a mission. Given a vector of resiliences,
$\mathbf{R}~ =~(R_1,R_2, \hdots, R_n),$ we define the overall cyber
resilence to be the
root-mean-square resilience of our mission.
\begin{equation*}
  \small
  R = ||\mathbf{R}||_2
   % = \left(\sum_{j=1}^n R_j^2 \right) ^{1/2}
    = \frac{1}{T-t_0}\left[\sum_{j=1}^n \left( \int_{t_0}^T F_j(\tau) \, d\tau \right)^2 \right]^{1/2}.
\end{equation*}
\normalsize
Because each mission accomplishment may have a different relative
importance, we may account for this by weighting the normal functionalities
${\Fnominal}_j$.  To enforce an overall mission resilience in the
range $[0,1]$, we renormalize to obtain normalized resilience $\mathcal{R}$.
\begin{equation*}
  \mathcal{R} = R \Big/\sqrt{\sum_{j=1}^n {\Fnominal}_j^2}.
\end{equation*}

\section{Case Study Analysis}

Figure~\ref{fig:fuel-efficiency} (top) shows the observed fuel efficiency plotted over the mission duration. The light blue line represents the observed fuel efficiency under the baseline scenario during which no attacks took place. The red line represents the fuel efficiency under two separate attack scenarios, with attacks occurring at $283~s$ and $619~s$, respectively. The effects of the attacks can be seen starting around $307~s$ and $669~s$, when the blue and red lines diverge. The loss in fuel efficiency, shaded orange, is the net effect of the attack on fuel efficiency.  The magnitude of the effect, calculated as the decrease in area under the curve (AUC) over the interval where the attack curves deviate from the baselines, was 10.76\% in the first scenario and 11.42\% in the second.

The experimental data were then analyzed using the novel mathematical framework detailed in a companion paper \cite{mmcr}. Briefly, the framework is used to express the functionality of a vehicle over mission time in terms of a parameterized change process. The key parameters are the magnitudes of the impact $\malware_0$ of malware and the impact $\bonware_0$ of % so-called
bonware (the totality of physical and cyber features that allow a system to resist and recover from cyber compromise). Since the effect of malware is not immediate, the effective onset of malware is modeled as an additional free parameter $t^m$.

The process model is governed by the differential equation $\frac{dF}{dt} = \left(F_N-F(t) \right) \bonware(t) - F(t) \malware(t) $,
where $\malware(t) = \malware_0 \, \Pi_{t^m,t^\star}(t)$ and $\Pi_{\alpha,\beta} = u(t-\alpha)-u(t-\beta)$ is the boxcar function, the difference of two unit step functions (resulting in a step from zero to one at time~$\alpha$ and from one to zero at time~$\beta$).  Bonware impact, $\bonware(t)=\bonware_0 u(t-t^\star)$ activates at the switching time.  Normative functionality, $\Fnominal$, is set to $1$ in this example. 

The process is applied to the functionality $F(t)$ of the vehicle over mission time. Functionality at time $t$ is defined as the ratio of a vehicular performance metric (here, fuel efficiency) at time $t$ in the attack scenario divided by the same metric at time $t$ in either the baseline scenario or the attack scenario, whichever is lower. Note that due to the high variability of fuel efficiency over short time intervals, the time series was first smoothed with a $72~s$ moving average window before taking the ratio (Figure~\ref{fig:fuel-efficiency}, top).

The parameters of the model can be estimated from data using standard model fitting tools. Our implementation relied on the estimation of a hidden Markov model as described in \cite{mmcr}. For the current data, we estimated $\malware_0 = 0.008$, $\bonware_0 = 0.048$, $t^m = 307~s$, and $t^\star=408~s$ for the first scenario and $\malware_0 = 0.008$, $\bonware_0 = 0.046$, $t^m = 669~s$, and $t^\star= 774~s\color{black} $ for the second. These model-based results allow us to quantify the relative impact of malware and bonware among different mission scenarios, vehicles, or attack and defense modalities.

Figure~\ref{fig:fuel-efficiency} (bottom) shows the functionality over time (solid light blue line) for the two case studies. Also shown is the functionality predicted by the three-parameter differential equation model (black line). The model captures the salient aspects of the data well.

\section{Discussion and Conclusions}\label{sec:conclusion}
The illustrative case study demonstrates that the experimental infrastructure
described generates data with sufficient fidelity to progress our model
for the quantitative measurement of cyber resilience.
Analysis of the data by subject matter experts confirm that it is similiar 
to data that would be generated by real-world systems.

The measurement of resilience is restricted to only
those \acp{KPP} that the system is able to
produce \cite{smith2022quantitative}.
In future work, components will be added to simulate different vehicles and
more \acp{KPP}.
This data may then be used to further refine our measurement methodology.
Also, similiar experiments will be conducted with an actual \ac{MGV}. These results will be compared to the simulated results to establish the
fidelity of the infrastructure and validity of the methodology.

\small
\bibliographystyle{IEEEtran}
\bibliography{bibliography.bib,references.bib,bib2.bib,IEEEabrv.bib}
\begin{acronym}[AutoCRATmm]
  \acro{ADF}[ADF]{Active Defense Framework}
  \acro{AUC}[AUC]{area under the curve}
  \acro{AutoCRAT}[AutoCRAT]{
    Automated Cyber Resilience Assessment Tool
  }
  \acro{AWD}[AWD]{all-wheel drive}
  \acro{BSFC}[BSFC]{brake-specific fuel consumption}
  \acro{CA}[CA]{Collaborative Agreement}
  \acro{CAN}[CAN]{Controller Area Network}
  \acro{CGW}[CGW]{central gateway}
  \acro{DIR}[DIR]{detect, isolate, and recover}
  \acro{DoD}[DoD]{Department of Defense}
  \acro{ECU}[ECU]{electronic control unit}
  \acro{IDE}[IDE]{integrated development environment}
  \acro{IP}[IP]{Internet Protocol}
  \acro{KPP}[KPP]{key performance parameter}
  \acro{LDI}[LDI]{livelihood diversification index}
  \acro{MEF}[MEF]{mission essential function}
  \acro{MGV}[MGV]{military ground vehicle}
  \acro{MQTT}[MQTT]{Message Queue Telemetry Transport}
  \acro{OBD2}[OBD2]{on-board diagnostics 2}
  \acro{PASTA}[PASTA]{%
    Toyota Portable Automotive Security Testbed with
    Adaptability
  }
  \acro{PGN}[PGN]{parameter group number}
  \acro{PID}[PID]{proportional-integral-derivative}
  \acro{POI}[POI]{point of interest}
  \acro{PSU/ARL}[PSU/ARL]{
    Penn State University Applied Research Laboratory
  }
  \acro{QMoCR}[QMoCR]{
    Quantitative Measurement of Cyber Resilience
  }
  \acro {RI}[RI]{resilience index}
  \acro {RPM}[RPM]{revolutions per minute}
  \acro {SAE}[SAE]{Society of Automotive Engineers}
  \acro {SLCAN}[SLCAN]{CAN over Serial}
  \acro {SPN}[SPN]{suspect parameter number}
  \acro {USB}[USB]{universal serial bus}
  \acro {CTIS}[CTIS]{Central Tire Inflation System}
  \acro{JLTV}[JLTV]{Joint Light Tactical Vehicle}
\end{acronym}

\end{document}